\documentclass[journal]{svmultln}

\usepackage{amsmath}
\usepackage{array}

 \usepackage[pdftex]{graphicx}

\usepackage{hyperref}

\usepackage{subfig}

\begin{document}
%

\title*{Instant Accident Reporting and Crowdsensed Road Condition Analytics for Smart Cities}

\titlerunning{Instant Accident Reporting and Road Condition Analytics}

\institute{Advanced Networks Research Lab\\
The University of Texas at Dallas\\
\href{https://anrlutdallas.github.io}{\textit{anrlutdallas.github.io}}
}


\author{Ashkan~Yousefpour, Caleb~Fung, Tam~Nguyen, David~Hong, Daniel~Zhang
	\thanks{This report is submitted to Communication Technology Changing the World Competition, Sponsored by IEEE Communication Society. Please refer to this link \textit{www.comsoc.org/communications-technology-changing-world} for more information.
}
}


%


\maketitle
\begin{abstract}
The following report contains information about a proposed technology by the authors, which consists of a device that sits inside of a vehicle and constantly monitors the car's information. It can determine speed, g-force, and location coordinates. Using these data, the device can detect a car crash or pothole on the road. The data collected from the car is forwarded to a server to for more in-depth analytics. If there is an accident, the server promptly contacts the emergency services with the location of the crash. Moreover, the pothole information is used for analytics of road conditions. 
\end{abstract}


%
\section*{General Description}
In recent years, the number of motor vehicle deaths has risen significantly. It is predicted that this trend will not slow down, especially with the growing population and the increased use of motor vehicles. Emergency vehicle response time plays a significant role in the event of these motor accidents. If emergency vehicles and response teams do not reach the scene of the accident in time, there could be devastating casualties. With the proposed technology, it is possible to reduce the response time of these emergency vehicles by eliminating the human factor in accident reporting, thus decreasing the amount of car accident fatalities. The proposed technology also brings to the city operators other benefits, such as Crowdsensed live pothole detection, which is intended to improve the quality of roads. This is a beneficial feature, since recently, research attention has also been directed toward roads condition monitoring (e.g. see \cite{basudan2017privacy}). 

The proposed technology consists of a device that sits inside of a vehicle and constantly monitors car information while driving. It can determine speed, g-force, and location. Using these data, the device can detect a car crash or pothole. If the device detects a crash, it promptly reports it to the server. The server then contacts the emergency services with the location of the crash. In addition, all this data are stored locally by the device onto an SD card so there are two copies of the data. This allows for the device to also function as a ``black box'' commonly found on commercial vehicles. The pothole information is used for analytics of road conditions. 

\section*{Technical Solution and Project Details}
The proposed accident reporting system can be divided into two modules/sub-systems. The first part is the Accident Detection Module (ADM) which is constantly monitoring acceleration and determines the moment a crash or pothole happens. It also continuously polls for GPS location and velocity. The second part is the Accident Reporting and Analysis Module (ARAM), which is a program and a web interface that has the online accidents reporting portal, analytics charts and the databases associated with the system. ARAM reports the accidents to Emergency organizations and calls the emergency contacts of the person(s) involved in accidents (see figure \ref{phone}). In the following subsections, we explain the two modules mentioned above.

\subsection*{Accident Detection Module (ADM)}
\begin{figure}[!t]
	\centering
	\subfloat[]{\includegraphics[width=0.3\linewidth]{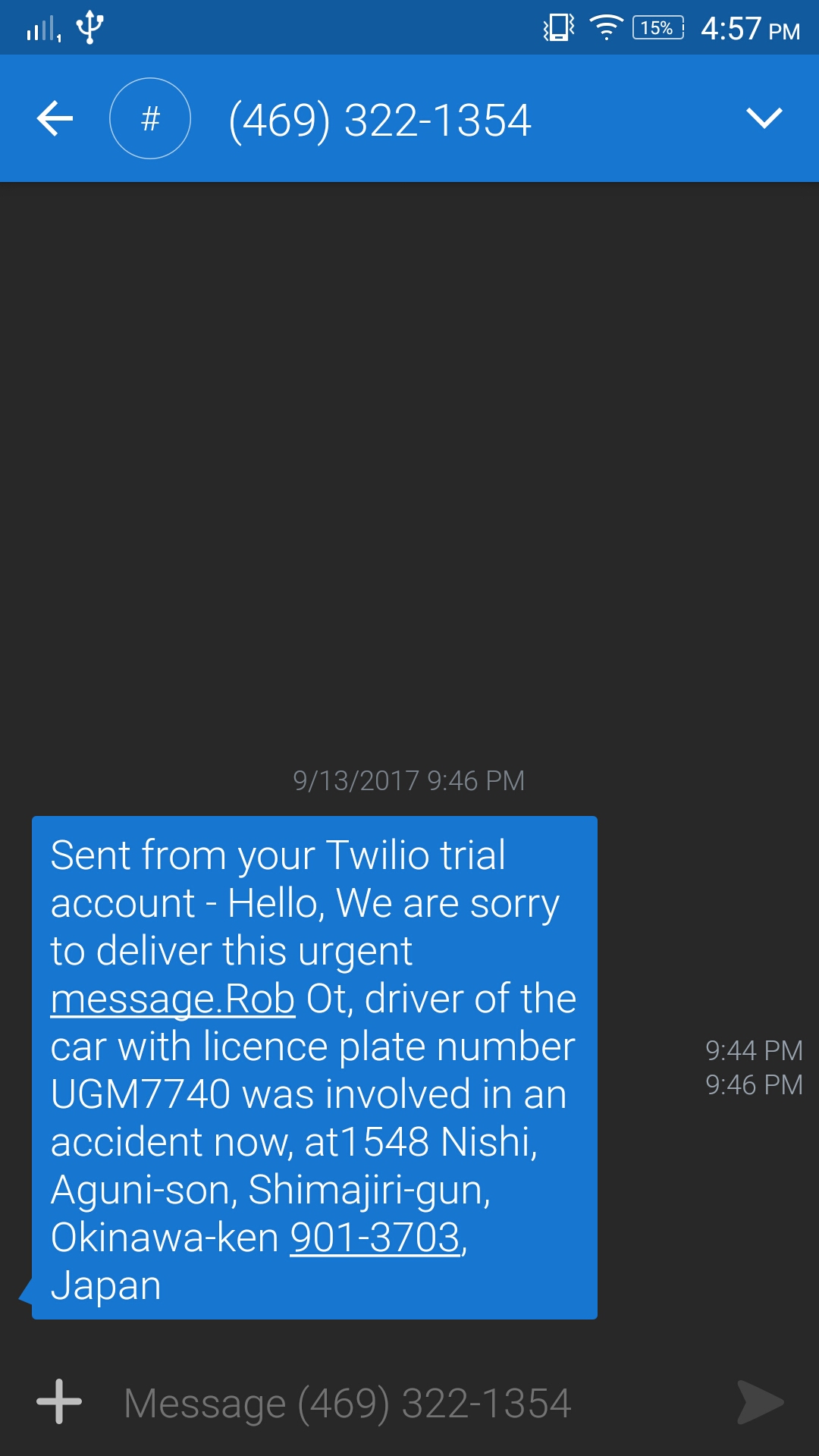}
		\label{phone}}
	\subfloat[]{\includegraphics[width=0.3\linewidth]{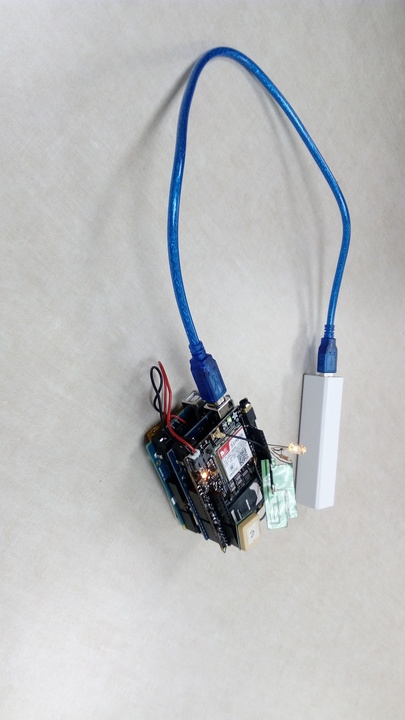}
		\label{hardware1}}
	\hfil
	\subfloat[]{\includegraphics[width=0.3\linewidth]{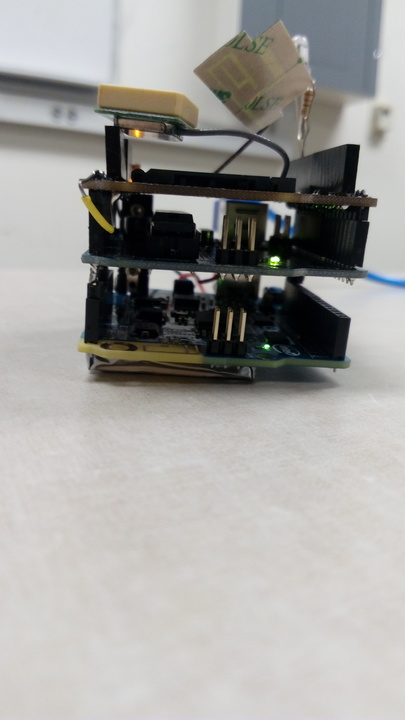}
		\label{hardware2}}
	\caption{(a) Emergency contact of the person who is involved in an accident gets a text message from the system. (b)(c) ADM device}	
\end{figure}

To make the ADM, we have used multiple hardware components: Arduino Uno, SIM808 Module, SIM card, GPS Antenna, GSM Antenna, Arduino 101 (Intel Curie), LEDs, 220 Ohm resistors. ADM is implemented on two separate microcontrollers. It is implemented on the Intel Quark microprocessor on the Intel Curie module, using an Arduino 101 as a prototyping board. The location and communication system is implemented on an ATmega328 microcontroller, with the Arduino Uno as the prototyping board and a SIM808 GSM and GPS module attached to it. The final ADM device is shown in figures \ref{hardware1} and \ref{hardware2}.  

The two microcontrollers act independently until there is a crash or pothole. The Arduino 101 sends an interrupt signal to the Arduino Uno and then promptly sends the crash data which includes the maximum axis of impact, the g-force enacted on that axis, and the total magnitude of all three axes as a percentage of the max value.

\begin{figure}
\centering
\includegraphics[width=0.8\linewidth]{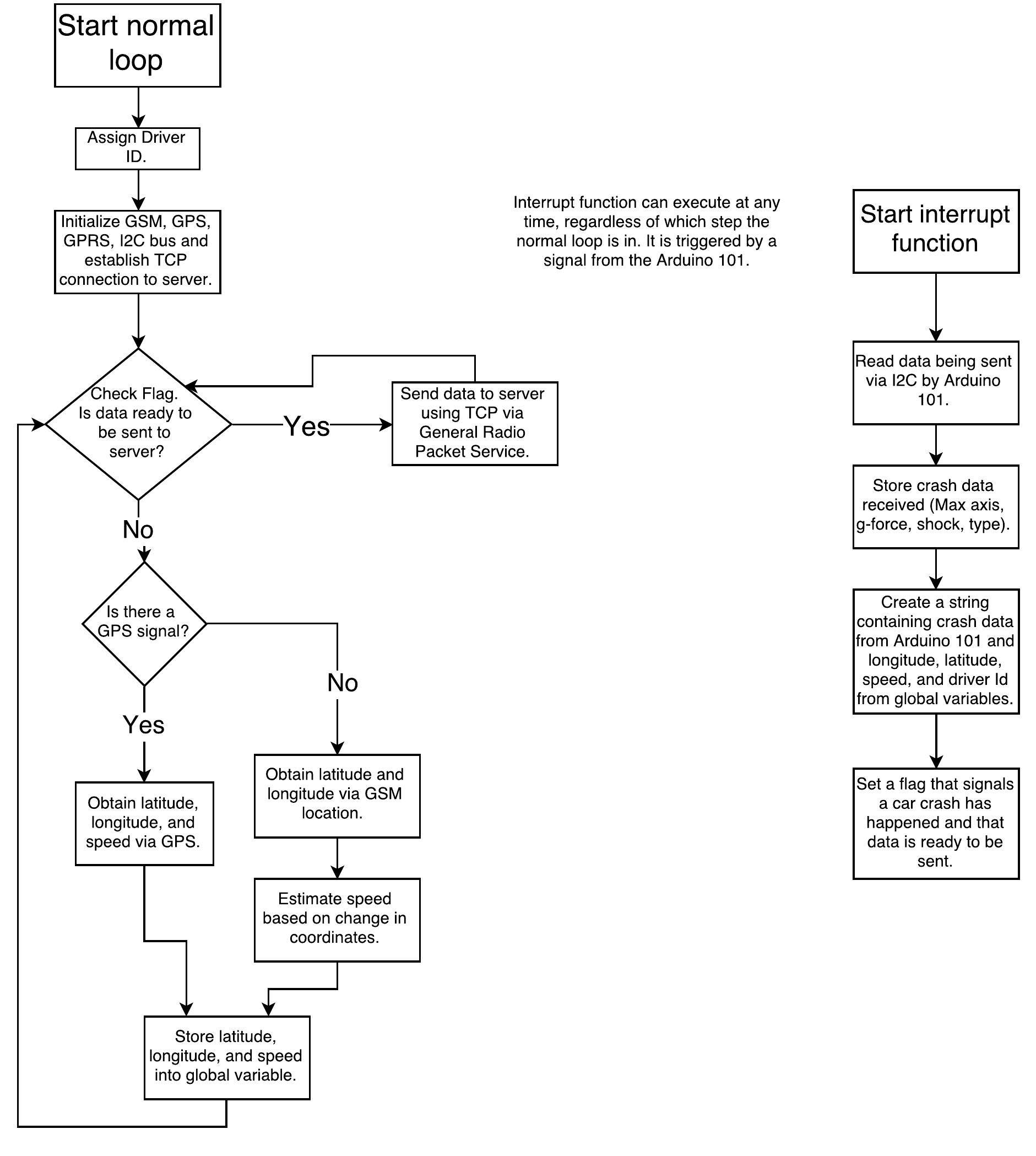}
\caption{The flowchart of Arduino Uno operation}
\label{uno}
\end{figure}

The Arduino Uno with the SIM808 module constantly polls for GPS latitude, longitude, and speed. The results are stored in respective variables and the variables are updated each time a new coordinate or speed is read. This is the main job of the Arduino Uno. While it does that, the Arduino Uno waits for an interrupt signal sent by the Arduino 101, which signifies a car crash. The operation of Arduino Uno is depicted in figure \ref{uno}.

The presence of a car accident can be detected by measuring a sharp change in acceleration, or impulse. This can be achieved using the accelerometer on the Intel Curie module on the Arduino 101. The accelerometer measures the acceleration along the x, y, and z axis in three-dimensional space. By dividing the measured values of acceleration in each axis by the acceleration due to gravity, the g-force enacted on each axis can be determined. The g-force value for all three axes are constantly polled and a car crash is declared when the g-force for any axis exceeds 12 G. The device determines whether the accident was a T-bone or a head-on collision based on whether the x or y axis of the accelerometer is negative or positive.

\begin{figure}
\centering
\includegraphics[width=0.8\linewidth]{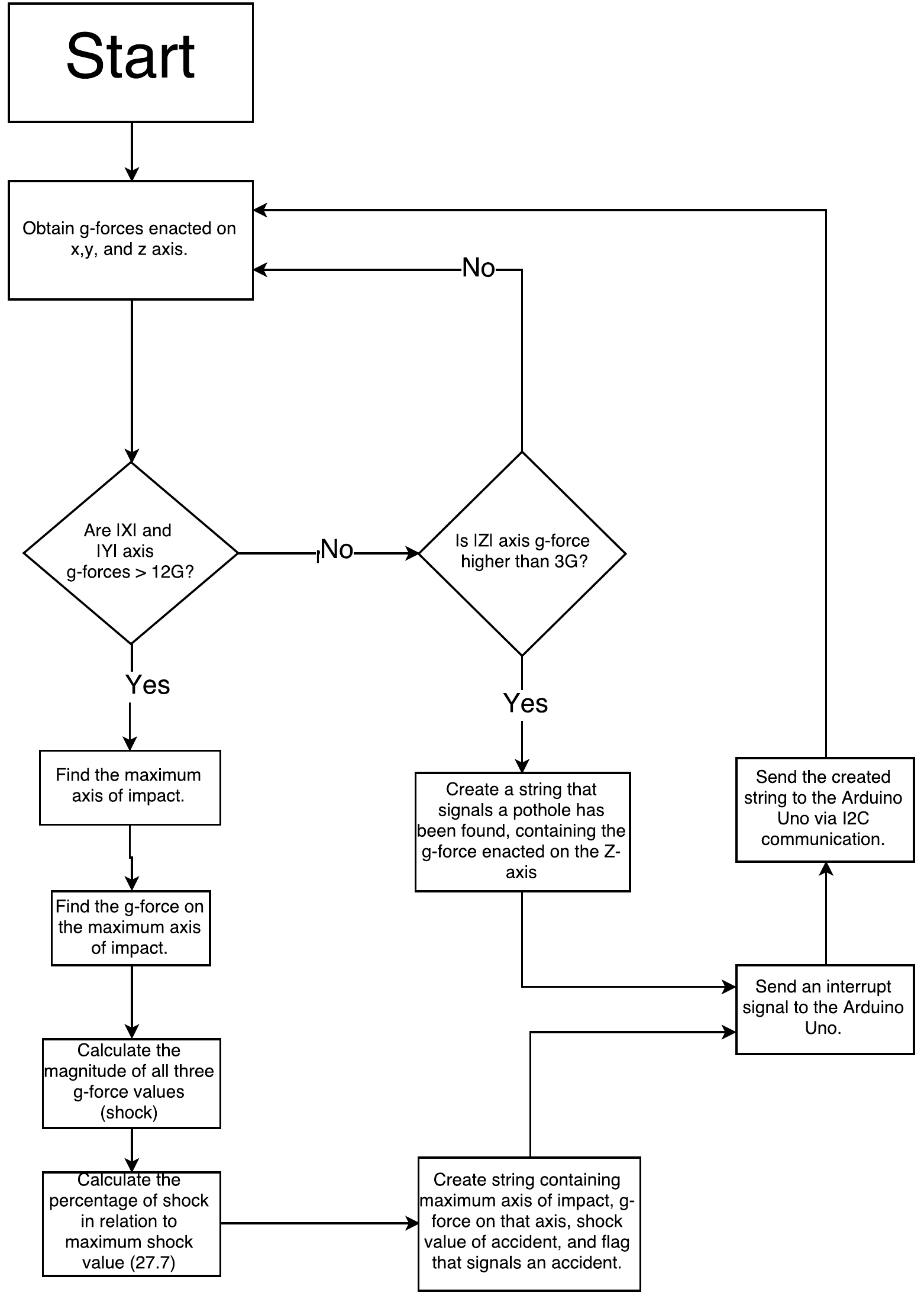}
\caption{The flowchart of Arduino 101 operation}
\label{101}
\end{figure}

When a car crash has been determined by the Arduino 101, an interrupt signal is sent to the Arduino Uno. The Arduino Uno then retrieves the axis with the maximum g-force enacted upon it, the g-force value on that axis, and the total magnitude of the shock as a percentage of the max. The Arduino Uno now has the GPS coordinates of the device, the last known speed before the crash, and the crash data. It compiles all of this information, appending a driverID to the end of all the data. The data are sent to the Accident Reporting and Analysis Module. The operation of Arduino 101 is shown in figure \ref{101}.

\subsection*{Accident Reporting and Analysis Module (ARAM)}
ARAM is implemented on Amazon AWS cloud IaaS, and acts as a server and a portal for the whole system. This module is implemented using the following technologies: JAVA, JavaScript, NodeJS, MySQL, JDBC, and Twilio, and it consists of three major components: server, events and drivers database, and web portal. The server is mainly responsible to receive, process and store the accident and pothole information that it receives from the ADMs of all cars. The events and drivers database contains information about drivers, their cars, and their emergency contacts; and events, such as accidents and pothole readings. The web portal visualizes the accidents and potholes (both live and over time), and also has analytics charts for the city operators. 

\begin{figure}
\centering
\includegraphics[width=\linewidth]{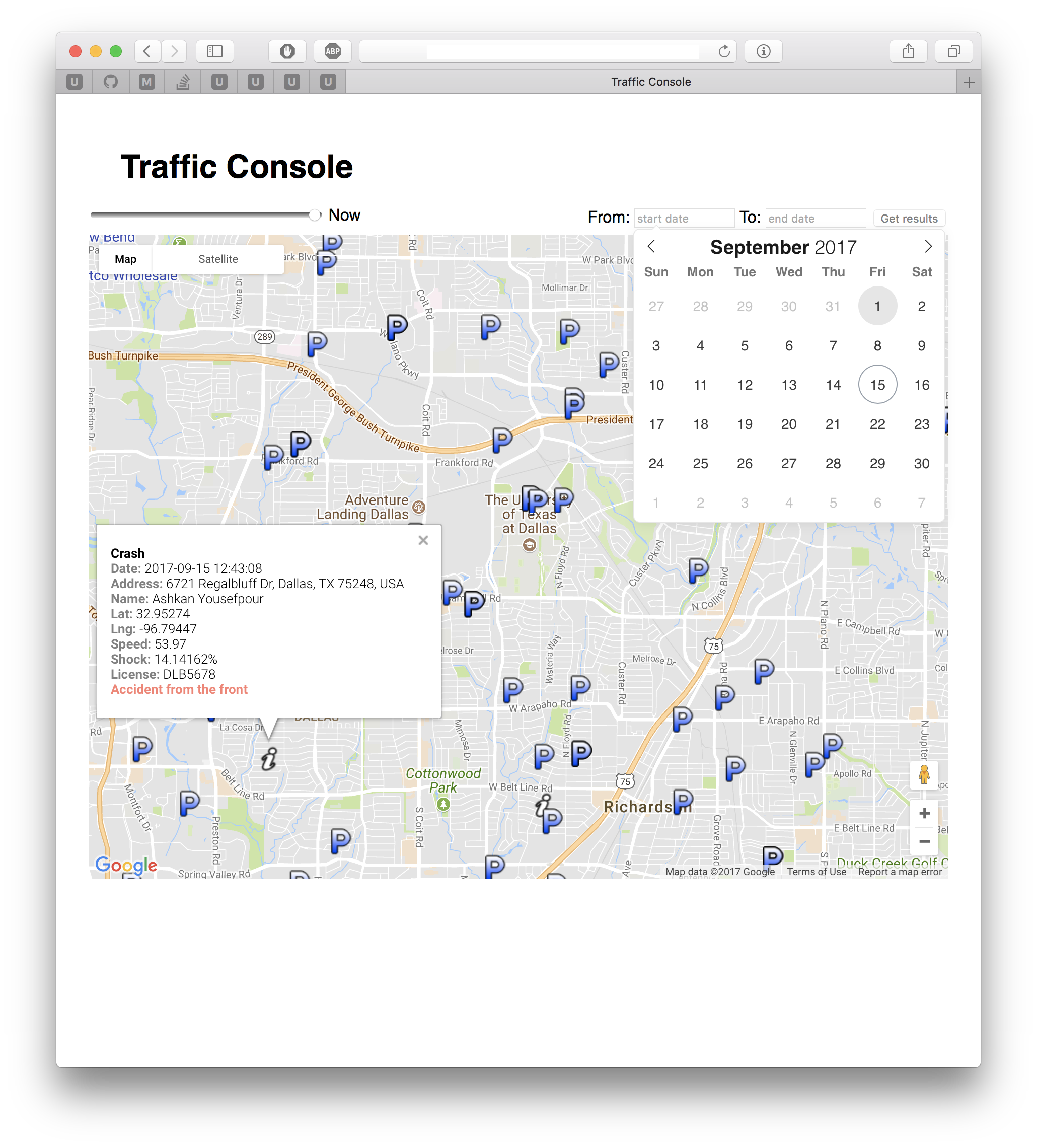}
\caption{Live Traffic Console (live accident and pothole monitoring)}
\label{console}
\end{figure}

The server is a multi-threaded program that listens and receives pothole and accident data from the installed ADMs on cars. When the server receives pothole information, it simply stores them in the events and drivers database, and also sends it to the web portal for live view. On the other hand, when the server receives accident reporting from an ADM, it first stores the information into the database, then it performs the emergency contacting. It does so by calling 911 and playing an auto-generated message, which consists of the name of the person(s) involved in the accident, the location of the accident, and the license plate of the car(s). The server also calls and texts the emergency contact of the person involved in the accident, with an appropriate message about the accident and its location. The server uses Google Reverse Geocoding library to get the human readable address of the accident from GPS latitude and longitude. 

\begin{figure}
\centering
\includegraphics[width=\linewidth]{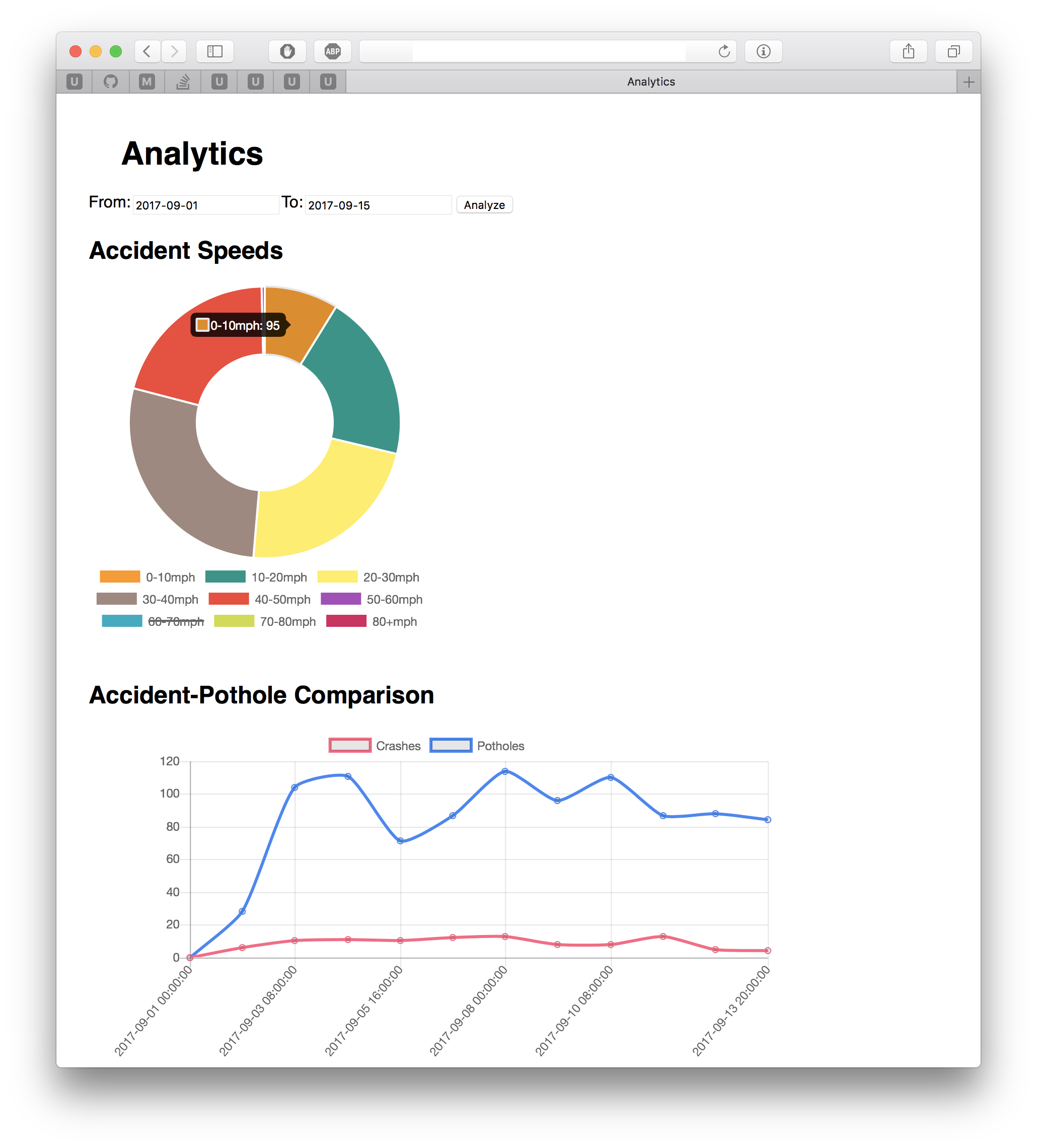}
\caption{Live Traffic Console (live accident and pothole monitoring)}
\label{analytics}
\end{figure}

The events table store the events generated by the system, i.e. potholes and accidents. It keeps track of events by storing latitude, longitude, speed, shock, type, and driver ID. The drivers table simply stores information about drivers, including their car, license plate of the car, and driver’s emergency contact information.

The web portal shows real-time data of traffic accidents and potholes using NodeJS. The web portal obtains data from the database and displays the data on a console webpage and analytics webpage (figure \ref{console}). In the console webpage, users can view past accidents and potholes for the past 24 hours or select a date range. In the analytics webpage (figure \ref{analytics}), users can view a comparison of accident speeds within a date range. They can also view a comparison of accidents to potholes over a time period that they input.

\section*{Social Impact}
According to the National Highway Traffic Safety Administration, in the United States, there were 35,092 reported traffic accident fatalities in the year 2015. To put in perspective, traffic accident fatalities in the year 2015 averages to 97 deaths per day from car accidents. The year 2015 saw the largest percent rise in car accident fatalities in 50 years at the rate 7.2\%. In 2016, the following year, motor vehicle deaths increased by 6\% \cite{NSC},\cite{transportDepartment}. Traffic accident fatalities are increasing at an alarming rate.

The proposed device aims to mitigate motor vehicle fatalities by alerting emergency services to the exact location of a vehicle accident the exact moment it happens. Studies have shown that the survival rate of motor accidents increases if the response time of emergency vehicles does not exceed five minutes \cite{blackwell2002response}. With such a short response time, every second is crucial, and that is the main motivation for the proposed technology.

With the current method of accident reporting, the reporter would first have to assess the severity of the accident. The reporter must then dial 911 and then relay the information to the dispatcher. This can take several minutes. For extremely severe crashes, the reporter would have to be a bystander not involved in the crash. This would further delay the time which the accident is reported since there may be no bystander at the moment of the accident. With our proposed scheme, this whole process can be reduced to an instant so that an emergency vehicle can be dispatched immediately. The system also contacts the emergency contacts of the person involved in the accident, so that they are informed of the accident. In many cases, when the emergency contact becomes present, many hospital-related procedures could be done faster, leading to a faster and more accurate medical care. 
\\
\section*{Implementation Status and Code}
 We have finished the implementation of all the components, including Arduino, server, database, and web portal. The code is available on a GitLab directory at  \href{https://gitlab.com/ANRL_UTD/IEEE-ComSoc-Competition-17}{\textit{https://gitlab.com/ANRL\_UTD/IEEE-ComSoc-Competition-17}}. The project is in working state. We have also tested the accident functionalities, such as calling emergency contacts, by generating accident reports from ADM. Nevertheless, installing the ADM in a car, and testing the system in an actual crash is not done yet, since it is difficult to make a car crash, only for the purpose of testing our product. 
 
 {\bf More info}: more information about this project, a demonstration video of the system, and guidance on the implementation details can be found at \href{https://anrlutdallas.github.io/resource/projects/accident-reporting.html}{\textit{https://anrlutdallas.github.io/resource/projects/accident-reporting.html}}

\bibliography{Master}{}
\bibliographystyle{ieeetr}

\end{document}